\documentclass[12pt,emtex]{article}
\usepackage{amsfonts}
\usepackage{amssymb}
\usepackage{amscd}
\usepackage{amsmath}
\newcommand{\BEQ}{\begin{equation}}
\newcommand{\EEQ}{\end{equation}}

\newtheorem{Lem}{Lemma}
\newtheorem{Prop}{Proposition}
\newtheorem{Cor}{Corollary}

\newtheorem{Claim}{Claim}

\newtheorem{Rem}{Remark}

\def\nn{\nonumber}

\def\bea{\begin{eqnarray}}
\def\eea{\end{eqnarray}}

\def\S{{\Sigma}}

\def\C{{\mathbb{ C}}}
\def\CC{{\mathbb{ C}}}
\def\RR{{\mathbb{ R}}}

\def\S2{$S^2$}

\newcommand{\bul}{{\bullet}}
\newcommand{\al}{{\alpha}}
\newcommand{\ga}{{\gamma}}
\newcommand{\h}{{\hbar}}
\newcommand{\pa}{{\partial}}

\def\l{\lambda}
\renewcommand{\gg}{{\mathfrak{g}}}

\def\one#1{#1^{\raise5pt\hbox{$\scriptstyle\!\!\!\!1$}}\,{}}
\def\two#1{#1^{\raise5pt\hbox{$\scriptstyle\!\!\!\!2$}}\,{}}

\begin{document}

\begin{titlepage}
\hfill ITEP-TH-28/04 \vskip 2.5cm

\centerline{\large \bf
Deformation quantization of submanifolds and reductions
} \centerline{\large \bf
via Duflo-Kirillov-Kontsevich map.
}
\vskip 1.0cm

\centerline{A. Chervov \footnote{E-mail:
chervov@itep.ru}
}

\vskip 1.0cm

\centerline{\sf Institute for Theoretical and Experimental Physics
\footnote{ITEP, 25 B. Cheremushkinskaya, Moscow, 117259, Russia.}}

\vskip 1.0cm

\centerline{L. Rybnikov \footnote{E-mail: leo\_ rybikov@mtu-net.ru}
}

\vskip 1.0cm

\centerline{\sf Moscow State University}

\vskip 2.0cm

\begin{abstract}

We propose  the following receipt to obtain the quantization
of the Poisson submanifold $N$ defined by the equations
$f_i=0$ (where $f_i$ are Casimirs)
from the known  quantization of the  manifold $M$:
one should consider factor algebra of the
quantized functions on $M$  by the images of $D(f_i)$,
where $D: Fun(M) \to Fun(M)\otimes \CC[\hbar]$
is Duflo-Kirillov-Kontsevich map.
We conjecture that this algebra is isomorphic
to quantization of $Fun(N)$ with Poisson structure inherited
from $M$.
Analogous conjecture concerning
the Hamiltonian reduction  saying that
"deformation quantization commutes with reduction"
is presented.
The conjectures are checked in the case of $S^2$
which can be quantized as a submanifold, as
a reduction and using recently found explicit star product.
It's shown that all the constructions coincide.

\end{abstract}

\vskip 1.0cm

\end{titlepage}

\tableofcontents


\section{Introduction}

\subsection{Quantization}

Consider a manifold $M$ with the Poisson bracket on it.
In \cite{Fronsdal} it was proposed to find the new
associative multiplication (usually called star product
and denoted by $f*g$)
 on $Fun(M)[[\hbar]]$ such that:
\bea \label{q} f*g=fg + \hbar(quantum~corrections)\mbox{  and  } f*g-g*f=i
\{f,g\} ~ mod ~ \hbar^2 \eea The algebra with the new multiplication
pretends to be the algebra of quantum observables associated with the
given classical algebra of observables which is $Fun(M)$. The problem was
ingeniously solved by Kontsevich in \cite{Kont}, (for the symplectic
manifolds it was done before in \cite{DeW}) who also obtained the
classification of the star products, which includes the following desired
result of the uniqueness: there is bijection between the star products up
to equivalence and Poisson brackets up to equivalence (see theorem in
section 1.3 in \cite{Kont}). The main property that equivalent star
products define isomorphic algebras on $Fun(M)[[\hbar]]$.

{\em
Hence despite  that for the given Poisson bracket one can
construct different star products satisfying \ref{q},
but the algebra corresponding
via Kontsevich's bijection
to the given Poisson bracket is unique up to isomorphism}.

{\bf Notation} we will denote such algebra by
$\widehat{Fun(M)}$.

Let us mention that from the physical perspective
the construction of the algebra of quantum observables
 is not the
full solution of the problem of quantization.
One also needs to define the Hilbert space
where such algebra acts unitary and irreducibly.
For the symplectic manifold it is believed
to be the only one such representation
(for the algebra $\widehat{Fun(M)}$ with $\hbar=1$).
This problem is not yet solved in full
generality,
but in some cases this can be done (see \cite{Kir-GQ,QKM1,BMS,Sc}).
In our paper one also finds the simplest example confirming this belief.

{\bf Notation} we will denote such Hilbert space
associated to the symplectic manifold $M$ as $H(M)$.

\subsection{Main conjectures}

The main aim of this paper is to propose and present some evidences
for the conjectures below.
Consider some Poisson manifold $M$ and some Casimir function
$f$ on this manifold (i.e. $f$ Poisson commute with any other
function). Then  submanifold $N:$ $f=Const$ inherits the Poisson
structure from $M$ (see section \ref{sect-inher} for explanations).
It is clear that:
\bea \label{clas-is}
Fun(N)=Fun(M)/(f-c)
\eea
as Poisson algebras, so it is natural
to try to quantize this isomorphism.
Let us denote by $D$ a  Duflo-Kirillov-Kontsevich
map from $Fun(M)\to \widehat{Fun(M)}$
(see section \ref{sect-DKK} for explanations).

{\bf Conjecture 1: }
There is isomorphism of algebras:
\bea
\widehat{Fun(M)}= \widehat{Fun(N)}/D(f-c)
\eea
One should possibly add some regularity property for
$f$ like $f=c$ is a smooth manifold,
(for several $f_i$ one should request
transversal intersection).

Let us mention that
due to results of Duflo, Kirillov, Kontsevich
$D(f-c)$ is Casimir in $\widehat{Fun(M)}$
i.e. $D(f-c)$  star product commutes with everything,
so ideal generated by it is both sided.

The conjecture 1 above can be generalized to the following
more general situation (called Hamiltonian reduction): consider
functions (called constraints)
$f_i\in Fun(M)$ such that they generate Poisson
closed ideal (such constraints are called  the
first class constraints following Dirac). Let us
denote by $I=I({f_i})$ the ideal generated by $f_i$.
Let $N=N({f_i})$ be Poisson normalizer of ideal $I$.
Let us consider Poisson factor algebra $N/I$, it is known
(at least for general $f_i$)
that it is algebra of functions on the quotient
of the manifold $f_i=0$ by the vector fields generated
by the Hamiltonian vector fields corresponding to $f_i$.
This manifold is called the Hamiltonian reduced manifold
by the constraints $f_i$ and denoted by $M//f_i$,
so there is the following isomorphism of Poisson
algebras:
\bea \label{clas-red}
Fun(M//f_i)=N/I.
\eea

The quantization of the Hamiltonian reduction
works as follows: denote by $\hat I$ left ideal
in $\widehat{Fun(M)}$ generated by $D(f_i)$,
by $ \hat N$ the right normalizer of it
in $\widehat{Fun(M)}$, so $\hat I$ is both
sided ideal in $\hat N$.

{\bf Conjecture 2: }
There is isomorphism of algebras:
\bea
\widehat{Fun(M//f_i)} = \hat N/ \hat I
\eea
One should most probably add some regularity properties for $f_i$
like transversal intersection and existence of
smooth quotient. Obviously conjecture 1 is particular
case of conjecture 2, because if  $f_i$ are
Casimirs, then  Hamiltonian vector fields corresponding
to them are zero and so $M//f_i$ is just the submanifold
$f_i=0$.

(The general scheme of the Hamiltonian reduction is due to Dirac \cite{Dirac},
our contribution is the remark that one
should use the map D to obtain the answer which is the quantization of the classically
reduced space).

This conjecture means that "deformation quantization commutes
with reduction" on the level of algebras of observables.
The same should be true for the Hilbert spaces associated
to the both manifolds i.e.

{\bf Conjecture 3:}
\bea
H(M//f_i)=\{ v\in H(M): D(f_i)v=0\}
\eea

In the case when $f_i$  are generators of some
compact Lie group  this conjecture is due to
Guillemin and Sternberg \cite{GS}.
(In this case there is no need to use the map $D$).
In  their work $H(M)$ was described in holomorphic
polarization. Their conjecture has been
proved recently (see \cite{S,V} for surveys).

Let us remark that our conjectures depends on some
auxiliary choices like the choice of concrete
generators in $f_i$ defining the ideal and
the choice of map $D$, which also not canonical,
we believe that the conjectures are true for
arbitrary choices mentioned above.

\subsection{Duflo-Kirillov-Kontsevich map}
\label{sect-DKK}

In \cite{Kont} M. Kontsevich proposed a universal method for deformation
quantization. Namely, for any Poisson bracket $\{\cdot,\cdot\}$ on a
manifold $M$ one can construct a star product on $Fun(M)$ such that
$a*b-b*a=i\hbar\{a,b\}\ (mod\ \hbar^2)$ for all $a,b\in Fun(M)$. The
gauge-isomorphism class of this star-product is defined canonically by the
gauge-isomorphism class of the Poisson bracket. This star-product
satisfies another very nice property: there is a natural mapping
\bea
D:Fun(M)\to\widehat{Fun(M)}
\eea
whose restriction to the Poisson center of
$Fun(\RR^n)$ gives an algebra isomorphism onto the center of
$\widehat{Fun(M)}$. We will call this map Duflo-Kirillov-Kontsevich map.

The construction of star-product in $\RR^n$  for arbitrary Poisson bracket
 is given by explicit, but very complicated
formula. The same can be said about the construction of the map $D$.
We postpone this definition to the appendix.
To our luck it was proved by Kontsevich  that in the case
when Poisson manifold is a vector space with linear Poisson
bracket (i.e. it is dual space to some Lie algebra with Kirillov's
bracket) it is true the following: quantization of such manifold
is isomorphic the  universal enveloping algebra and
the map $D$ coincides  with the rather explicit
map called Duflo-Kirillov  map.
In this paper we will consider only such Poisson manifolds.
So here we will recall the definition of the map $D$ in this situation.

In the case when Poisson manifold is $\RR^{2n}$ with the
standard Poisson bracket $\{p_k,q_j\}=\delta_{kj}$ then
{\em the map $D$ is just the symmetrization}.
\bea
D(ab)=\frac{1}{2!}(a*b+b*a), \
D(a_1a_2...a_n)=\frac{1}{n!}(\sum_{\sigma \in S_n}
a_{\sigma(1)}*a_{\sigma(2)}*...*a_{\sigma(n)})
\eea
where $a,b,a_{i}$ are
any linear combinations of $p_k,q_j$. It is easy to check that the map $D$
gives an $sp_{2n}$-module isomorphism between $Fun(\RR^{2n})$ and
$\widehat{Fun(\RR^{2n})}$.

The more general case is the following:
 Poisson manifold is a $\gg^*$
with the Kirillov's bracket, where  $\gg^*$ is the dual space  of a
finite-dimensional Lie algebra.  The algebra $Fun(\gg^*)$ is the symmetric
algebra $S(\gg)$ and $\widehat{Fun(\gg^*)}$ is isomorphic to  the universal enveloping
algebra $U(\gg)$. To define the Duflo-Kirillov-Kontsevich map in this case
we introduce some notations.
\begin{enumerate}
\item Let $Tr_{2k}$ be invariant polynomials on $\gg$,
$x\mapsto Tr_{\gg}(ad x)^{2k}$, considered as differential operators on
$\gg^*$ with constant coefficients.
\item Let $a_{2k}$ bee the sequence of real numbers, such that \bea
\sum\limits_{k\ge0}a_{2k}t^{2k} =
\frac{1}{2}Log\frac{e^{\frac{t}{2}}-e^{-\frac{t}{2}}}{t}. \eea
\item Let $\sigma:S(\gg)\to U(\gg)$ be the symmetrization map.
\end{enumerate}
The Duflo-Kirillov-Kontsevich map $D:S(\gg)\to U(\gg)$ is given by the
formula (see \cite{Kont}, section~8.3)
\bea
s\mapsto\sigma(e^{\sum\limits_{k\ge 0}(i\hbar)^{2k}a_{2k}Tr_{2k}}s).
\eea
Note that in the case of nilpotent $\gg$ we have $Tr_{2k}=0$ for all $k$,
and hence $D=\sigma$. The different proofs that this formula
gives the isomorphisms of centers of $S(\gg)$ and  $U(\gg)$
can be found in
\cite{Duflo} and \cite{Ginzburg}.

\subsection{Plan of the paper}

The main text of this paper is devoted to the successful check of our
conjectures in the first nontrivial example of the sphere \S2 with standard
$SO(3)$ invariant symplectic form. Sphere \S2 with this symplecitc form can
be obtained in two different ways first as coadjoint orbit for $SO(3)$ i.e.
as a submanifold in $so(3)$ second as the Hamiltonian reduction of $\RR^4$.
Both constructions  can be quantized according
to our receipts and we see that the results coincide.
This is done in section \ref{sect-q-s2}. Conjecture 3
is also true in this example see section \ref{sect-hil-sp-red}.

The third thing - we compare the quantizations
above with the explicit star product construction
recently found in \cite{jap}.
And also we find the complete agreement.
This is done in section \ref{sec-ex-star}.

{\bf Acknowledgements.} This paper originates from
the discussion on the seminar in ITEP.
We are indebted to our friends, participants
of this seminar for the discussions and criticism:
N. Amburg, S. Galkin,
S. Gorchinsky, A. Gorodentcev, S. Loktev, M. Mazo,
 G. Sharygin, K. Shramov, D. Talalaev. We are also deeply
indebted to V. Dolgushev and A. Karabegov for the explanations and
to M. Vergne, J. Rawnsley and A. Karabegov for sending us their papers.
The  work of the authors  has been partially supported by the grants:
AC by the RFBR grant 04-01-00702, LR by the CRDF grant RM1-2543-MO-03.


\section{Quantization of  \S2 as a submanifold} \label{sect-q-s2}

In this section we consider \S2 as a submanifold
in $\RR^3=so(3)$ and we quantize it by the receipt
of the conjecture 1 (see claim \ref{claim-orbit} in section
\ref{sect-duflo}).
We also show that our considerations completely confirm
the belief that for the symplectic manifold there is only
one irreducible and unitary representation of
the algebra $\widehat{Fun(M)}$ (see section \ref{sssect-hilb-space0}).

\subsection{Inheriting the symplectic structure on \S2 from $\RR^3=so(3)$.}
\label{sect-inher}

In this subsection we will explain how to inherit
the  symplectic structure from Poisson bracket in
$\RR^3$ and calculate the volume.
(It does not  coincide with Eucledian volume $4\pi R^2$,
but is given by $2\pi R$).

Consider the sphere in $\RR^3$ given by the equation
$x_1^2+x_2^2+x_3^2=R^2$.

The space $\RR^3$ can be identified with the Lie algebra $so(3)$
(more precisely  with its dual space $so(3)^*$ but in
the case of semisimple Lie algebras like so(3) one can identify
$so(3)$ and $so(3)^*$ with the help of Killing form),
so it can be endowed with the Poisson bracket $[x_i,x_j]=2\epsilon_{ijk}x_k$,
where $\epsilon_{ijk}$ is totally antisymmetric tensor.

The element $C=x_1^2+x_2^2+x_3^2$ is Casimir element for this bracket, so
the sphere $x_1^2+x_2^2+x_3^2=R^2$ can be endowed with the Poisson bracket.
This is the general trivial fact that: if any element $C$ is Casimir for any
Poisson bracket on any manifold $M$, then the submanifold $N$: $C=Const$,
inherits the Poisson bracket from $M$. Which goes as follows:
the functions on $N$ are factor by the ideal generated by $C$
of functions on $M$, but the ideal $I$  generated by the Casimir $C$
is ideal with respect to the Poisson bracket, i.e.
for $ f\in I, g\in Fun(M)$  holds $\{ f, g \} \in I$.
Hence the Poisson bracket can pushed down to the
factor algebra $Fun(M)/I=Fun(N)$.

Let us mention that the only property needed for
restricting the Poisson bracket to the submanifold is the property
that ideal $I$ is Poisson ideal.
Geometrically this can be reformulated as bivector $\pi$
is tangent to the submanifold N. (Polyvector is tangent
to some submanifold iff it can be presented as sum of products
of tangent to this submanifold vectors).

Obviously Poisson bracket on \S2 is nondegenerate,
so we obtain symplectic form on \S2.
It is easy to see that \S2 is coadjoint orbit for
$so(3)$, and the symplectic structure above is
Kirillov's symplectic structure on the coadjoint orbit.
It is obiously $so(3)$ invariant.

{\Lem \label{volume}
volume of \S2 with respect to this symplectic form
is $2\pi R$.}

{\bf Proof} at the upper half sphere one can consider
$x_1,x_2$ as local coordinates and the Poisson bracket
is given by $\{ x_1, x_2 \}= 2x_3=2\sqrt{R^2-x_1^2-x_2^2}$,
so the symplectic form is given by
$\omega=\frac{1}{2\sqrt{R^2-x_1^2-x_2^2}} dx_1\wedge dx_2$.
So the volume of the semisphere can be calculated
as
\bea
\int_{x_1^2+x_2^2 < R^2} \frac{1}{2\sqrt{R^2-x_1^2-x_2^2}} dx_1\wedge dx_2
=\int_{0<r<R,0<\phi<2\pi}\frac{1}{2\sqrt{R^2-r^2}}rdrd\phi
=\nn\\
=\int_{0<r<R,0<\phi<2\pi}\frac{1}{4\sqrt{R^2-r^2}}dr^2d\phi
=2\pi \int_{0<r<R}\frac{R}{4 \sqrt{1-\frac{r^2}{R^2}}}d\frac{r^2}{R^2}
=\nn\\
=2\pi  \int_{0<u<1}\frac{R}{4 \sqrt{1-u}}du
=-2\pi R \frac{1}{2} \sqrt{1-u}_{0}^{1}=\pi R
\eea
Now recalling that it is volume of semisphere
we multiply it by two and obtain the volume of sphere
is $2\pi R$.
$\Box$

\subsection{Quantization and Duflo-Kirillov map.}\label{sect-duflo}

The algebra of functions on $\RR^3$ with respect
to star product corresponding to
Poisson bracket $\{x_i,x_j \}=2\epsilon_{ijk}x_k$ is isomorphic
to $U(so(3))$.
(This is true for any Lie algebra
see for example \cite{Kont} section 8.3.1).
In this section we will never use the star-product,
but we will work with $U(so(3))$. Let us denote the multiplication
in $U(so(3))$ by $\odot$.
So $[x_i,x_j]_{\odot}=2i\hbar\epsilon_{ijk}x_k$.

{\it Later on we put $\hbar=1$ for simplicity.}

{\Prop The image of Casimir element $C=\sum_i x_i^2$ under Duflo-Kirillov
isomorphism is $D(C)=\sum_i x_i \odot x_i + 1 \in U(so(3))$}.

{ \bf Proof:} Explicit computation shows that $a_2=\frac{1}{48}$ and
$Tr_2=-8\sum_i \partial_i^2$. Hence we have
\bea D(C)=\sigma(C+1) = \sum_i
x_i\odot x_i + 1.
\eea
$\Box$

As a corollary of this proposition
we obtain that modula the conjecture 1 the following theorem
is obtained. (We call "claim"   because it is proved here
only modula conjecture 1, later we will prove it by explicit
star product construction, so it will be really the theorem):

{\Claim  \label{claim-orbit}
 Quantization of \S2 with standard SO(3) invariant
symplectic form of the volume $R$  (i.e. the algebra $\widehat{Fun(S^2)}$
with $\hbar=1$) is isomorphic to the algebra
\bea
U(so(3))/(\sum_i x_i\odot x_i + 1=R^2),
\eea
where $x_i$ are generators of so(3) obeying the relations:
$[x_i,x_j]_{\odot }=2i\epsilon_{ijk}x_k$}.

So we have described the quantization of \S2 as a submanifold
in $\RR^3$.

\subsection{Hilbert space  from representation theoretic
point of view.}
\label{sssect-hilb-space0}

Now let us describe the (unique) finite-dimensional representation of the
algebra $U(so(3))/ (D(C)=R^2)$.

Recall that the isomorphism of $sl(2)$ and $so(3)$ is given by the formulas
\bea
h=x_1,\
e=\frac{1}{2}(x_2 +i x_3),\  f=\frac{1}{2}(x_2 - i x_3).
\eea
The commutator relations os $sl(2)$ are
standard $[e,f]_{\odot} =h, [h,e]_{\odot} =2e, [h,f]_{\odot} =-2f$.
The Casimir
element $D(C)\in U(so(3))=U(sl(2))$ can be rewritten as
\bea
D(C)=\sum_i x_i \odot x_i + 1= 4e\odot f+h\odot h-2h+1=4f\odot e+h\odot h+2h+1.
\eea
Let $V_{\l}$ is
the irreducible representation of the Lie algebra $so(3)=sl(2)$ with the
highest weight vector $|0>$ and weight $\l=R-1$, i.e. $ h|0> = \l |0>,\ e
|0>=0 $. The Casimir operator $D(C)=\sum_i x_i \odot x_i + 1=
4e\odot f+h\odot h-2h+1=4f\odot e+h\odot h +2h+1$ acts on it as scalar operator on $V_{\l}$ and
the scalar can be easily computed
\bea
\label{cas} D(C)|0>=
(4f\odot e+h\odot h +2h+1) |0>= (\l^2+2\l+1) |0> = R^2 |0>.
\eea

So we come to the following lemma:

{\Lem \label{lem-rep-th-hilb}
We see from representation theoretic point of view
that the belief that the algebra of quantized functions
has the only representation in the Hilbert space finds complete
confirmation. The only representation is $V_{R-1}$.
Its dimension is equal to $R$.
Other representation of $sl(2)$ should be dropped
out because either Casimir will act by the irrelevant
constant or because they cannot be made unitary (like Verma modules).
}


\subsection{Hilbert space  from  geometric quantization}
\label{sssect-hilb-space1}

According to general optimistic belief the deformation quantization of the
algebra of functions (with $\hbar=1$)
 on the symplectic manifold $M$ has unique irreducible
unitary (i.e. real-valued functions acts
as self-adjoint operators)
representation in the Hilbert space. (For the Poisson manifold
the representations are related to the symplectic leaves).
Moreover the dimension of the such representation
is expected to be
 given by the formula $\int_M exp(\omega) \hat A(M)$, where
$\hat A(M)$ is A-genus of the manifold $M$.
This is predicted by the geometric quantization
with half-forms and by Fedosov's index theorem \cite{Fed}
(one usually requests $\omega$ to be integer
2-form on $M$, but possibly for non integer
2-forms all the same can be done making from $\widehat{Fun(M)}$
von Neumann algebra and using
von Neumann's  fractional dimension).
(If $K$ is trivial and in some other cases
 this coincides with the naive prediction
of physicists that the dimension
of the Hilbert space is 1/n!(symplectic volume),
usually this is said in textbooks
as one quantum state takes
$ \frac{\prod dp_i \prod dq_i}{(2\pi\hbar)^n}$
of the phase space (see for example section 48 in \cite{LL})).
If the form $\omega$ is Kaehler and sufficiently positive
form then
Hilbert space can be realized as the space of
holomorphic sections $H^{0}(L\otimes \sqrt{(K)})$,
where $L$ is such line bundle that: $c_1(L)=\omega$,
$K$ canonical line bundle (the line bundle of
holomorphic exterior forms of highest degree).
The line bundle $\sqrt{(K)}$ is such bundle
that $\sqrt{(K)}\otimes \sqrt{(K)}=K$, it exists
if $w_2(M)=0$ and unique if $M$ is simply connected.
The sections of such bundle are called half-forms.
Note that by the Riemann-Roch  and vanishing theorems
$dim H^0(L\otimes \sqrt{K})= \int_M exp(\omega-\frac{1}{2}c_1(M))Todd(M)
=\int_M exp(\omega) \hat A(M)$, due to the equality
$ exp(-\frac{1}{2}c_1(M))Todd(M) =  \hat A(M)$.

One receipt which is due to Kostant and Souriau how to construct the representation
of the algebra of functions with deformed
product in the space of sections of some line bundle
is called geometric quantization (see \cite{Kir-GQ} for survey).
(Let us mention that it was developed before
deformation quantization,
and there is some misunderstanding that sometimes
people insist on the exact equality $i\{f,h\}=[f,g]$
in the geometric quantization approach. This is not really true.
This is true
only for consideration of representation on the non-polarized sections,
but when one needs to find the representation in the
Hilbert space i.e. in the space of polarized sections - this commutation
relation does not hold).
Though it was never realized in full generality,
it is known to work in the case of semisimple orbits
of semisimple Lie algebras.
(Another receipt is the so-called Berezin-Toepltiz
quantization which succeeds in the case of compact Kaehler
manifolds \cite{QKM1,BMS,Sc}.)

Turning from the generalities to our concrete example
of \S2=$\CC P^1$ we see that geometric quantization
predicts that the algebra of functions
with the star product should have
the irreducible unitary representation  realized in
the sections of line bundle
$L\otimes \sqrt{K}$, where  $L= {\mathcal{O}}(R)$,
on $\CC P^1$ it is well-known that $K=O(-1)$
and so $L\otimes \sqrt{K} = {\mathcal{O}}(R-1)$.
Hence the dimension of the Hilbert space
is $dim H^0(L\otimes \sqrt{K})$ and it is equal to:
\bea \label{dim}
dim H(S^2)=
\int_{S^2} exp(\frac{1}{2\pi} \omega )exp(-\frac{1}{2}c_1(S^2))
\hat A (S^2)= \int_{S^2} (\frac{1}{2\pi} \omega ) =
 R.
\eea

{\Rem ~} We see that dimension of the Hilbert space
in this example coincides with the symplectic volume
up to $2\pi$.

{\Cor We see the complete agreement for the dimension of
Hilbert space prescribed from the representation theoretic
of view (see Lemma \ref{lem-rep-th-hilb}) and
from the point of view of geometric quantization
with half-forms}.


\section{Quantization of \S2 by Hamiltonian reduction.}

In this section we  recall the Hamiltonian
reduction procedure and we show how to
obtain \S2 as a reduction of $\RR^4$,
and proceed with quantization of reduction
by the receipt of the conjecture 2
(see claim \ref{claim-red} in section
\ref{sect-q-red}).
As an evidence for
our conjectures we
show that the result is the same as in the
previous section
(see corollary
\ref{cor-iso} in section \ref{sect-q-red}).
We also confirm  the conjecture 3 describing the Hilbert space from the point
of view of the reduction.

The procedure
of Hamiltonian reduction is due to Dirac
\cite{Dirac} (see \cite{Merkulov} for
recent short exposition and  very nice
remark that non reduced constraints like
$x^n=0$ leads to appearance of matrix
degrees of freedom, which was possibly
motivated by string theoretists belief
that coincident D-branes leads to
appearance of  U(n) gauge group
as "brane volume" theory).
The geometric sense of the Hamiltonian
reduction in the case of arbitrary symplectic
manifolds was explained to mathematicians
in \cite{MW}.



\subsection{Classical Hamiltonian reduction of $\RR^4$ by
$\frac{1}{2}(p_1^2+q_1^2+p_2^2+q_2^2)$.} \label{sect-cl-red}

The procedure of hamiltonian reduction has been briefly described in
the introduction, we will follow the described scheme.

The symplectic structure on \S2 considered above can be obtained as a
Hamiltonian reduction of the constant symplectic structure on $\RR^4=\C^2$.
Namely, let $p_1,p_2,q_1,q_2$  be coordinates on $\RR^4$
with  the standard Poisson bracket
 (i.e. $\{p_i,q_j\}=\delta_{ij}$), and let
$z_1=\frac{1}{\sqrt{2}}(q_1+ip_1), z_2=\frac{1}{\sqrt{2}}(q_2+ip_2)$, then
$\{z_1, \bar z_1\}=i, \{ z_2, \bar z_2\}=i$.
Let us consider the constraint, which is Hamiltonian for the
harmonic oscillator:
\bea
E=\frac{1}{2}(p_1^2+q_1^2+p_2^2+q_2^2)=(\bar z_1 z_1+ \bar z_2 z_2).
\eea

{\Lem
Let $N$ be the
commutant in $Fun(\RR^4)$ of the element $E$ with respect to the Poisson
bracket. The algebra $N$ is generated by:
\bea \label{x-i}
E,\ x_1=\frac{1}{2}(p_1^2+q_1^2-p_2^2-q_2^2)=
(z_1\bar z_1-z_2\bar z_2),\nn\\
x_2 = (q_1q_2+p_1p_2)= {2} Re (z_1\bar z_2),\
x_3 = (p_1q_2-q_1p_2)= {2} Im (z_1\bar z_2)
\eea
The elements $x_i$ satisfy the relations:
$\{ x_i, x_k \}=2\epsilon_{ijk}x_k$,
which are so(3) relations.
}

{\bf Proof.} Clear.

Let us recall that we have defined Casimir element in so(3) as
$C=\sum_i x_i^2$

{\Lem $C= E^2$.}

{\bf Proof.} $C=x_1^2+(x_2+ix_3)(x_2-ix_3)=
(|z_1|^2-|z_2|^2)^2+4|z_1|^2|z_2|^2
= (|z_1|^2+|z_2|^2)^2$.
$\Box$

The element $E$ is central in $N$, hence the Poisson bracket on the
algebra $S=N/(E=R)$, where $R$ is a constant, is well-defined.

{\Cor
The classical Hamiltonian reduction of $\RR^4$ by the constraint
$E-R= \frac{1}{2}(p_1^2+q_1^2+p_2^2+q_2^2)-R$ is sphere $S^2$ of
symplectic volume $2\pi R$.
On the level of functions this mean that:
$ N/(E=R)$ is isomorphic to $Fun(S^2)$ as Poisson algebra
and isomorphism is given by formulas \ref{x-i}.
}

{\bf Proof:} the calculation of volume follows
from proposition \ref{volume}, the other things are clear.


\subsection{Quantum Hamiltonian reduction of $\RR^4$ by
$\frac{1}{2}(p_1^2+q_1^2+p_2^2+q_2^2)$.} \label{sect-q-red}

As we have already mentioned in the introduction
the quantization of $\RR^{2n}$ can be explicitly
described by the Moyal formula \cite{Moyal}:
\bea r*s=
(e^{i\hbar\frac{1}{2}\sum\limits_{i=1,2}\partial_{p_i}\partial_{\tilde{q_i}}
-\partial_{q_i}\partial_{\tilde{p_i}}}
r(p,q)s(\tilde{p},\tilde{q})|_{p_i=\tilde{p_i},\ q_i=\tilde{p_i}})
\eea

Let us put $\hbar=1$.

All commutators in this section are with respect to the Moyal's product.

It's obviously true that
$[p_i,q_j]=i\delta_{ij}$ (hence
$z_i$ satisfy the relations: $[z_i, \bar z_i]=1$).

Recall that the Duflo-Kirillov-Kontsevich map
in this case is given just by the symmetrization:
\bea
D(a_1...a_n)=\frac{1}{n!}\sum_{\sigma\in S_n}
a_{\sigma(1)}*...*a_{\sigma(n)},
\eea
where $a_k$ is any linear combination of $p_i,q_j$.

Thus, on the quantum level we have \bea \label{hat-e} D(E)=\hat E=
\frac{1}{2}(p_1^2+q_1^2+p_2^2+q_2^2)=
\frac{1}{2}(p_1*p_1+q_1*q_1+p_2*p_2+q_2*q_2)=\nn\\
=
\frac{1}{2}\sum\limits_{i=1,2}z_i*\bar z_i+\bar z_i *z_i.
\eea

{\Rem ~} so let us mention that in this case
if one works in generators
$p_i,q_j$ then there is no need
to use the symmetrization due to
$E=D(E)$, but working in generators
$z_i,\bar z_i$ really shows that
symmetrization is really
essential due to
$D(E)= \frac{1}{2}\sum\limits_{i=1,2}z_i*\bar z_i+\bar z_i *z_i
\neq \sum\limits_{i=1,2}z_i*\bar z_i \neq
\sum\limits_{i=1,2}z_i*\bar z_i $.

{\Lem For any $s\in Fun(\RR^4)$ we have
$[D(E),D(s)]=D (\{E,s\})$.}

{\bf Proof.} Indeed, let $s$ be homogeneous of degree $m_i$ with respect
to $z_i$ and of degree $n_i$ with respect to $\bar z_i$. Then
$[D(E),D(s)]=(n_1+n_2-m_1-m_2)D( s) =D((n_1+n_2-m_1-m_2)s)=D(\{E,s\})$.
$\Box$

{\Cor \label{cor-commutant}
 Denote by $\hat{N}$ the commutant of $D(E)$. This algebra is generated
by
\bea
D(E),\ x_1=\frac{1}{2}(p_1^2+q_1^2-p_2^2-q_2^2)=
\frac{1}{2} (z_1*\bar z_1-z_2 *\bar z_2+\bar z_1* z_1-\bar z_2 * z_2),\nn\\
x_2 = (q_1*q_2+p_1*p_2)= (q_1q_2+p_1p_2) = {2} Re (z_1*\bar z_2)
={2} Re (z_1\bar z_2),\nn\\
x_3 = (p_1*q_2-q_1*p_2)= (p_1q_2-q_1p_2)=  {2} Im (z_1*\bar z_2)
=  {2} Im (z_1\bar z_2)
\eea
}


{\Lem \label{lem-com-rel}
Elements $x_i$ satisfy the relations
$[x_i,x_j]=2i\epsilon_{ijk}x_k$, and  hence
elements ${h}=x_1, e =\frac{1}{2}(x_2+i x_3)=z_1\bar z_2,
{f}=\frac{1}{2}(x_2-i x_3)= \bar z_1 z_2 $ satisfy the sl(2)
relations: $[e,f]=h, [h,e]=2e, [h,f]=-2f$.
}



Let us recall that according to conjecture 2
about the Hamiltonian reduction the quantizations of functions on \S2
should be $\hat{N}/ (D(E)=R)$. So we see that:

{\Claim  \label{claim-red}
 Quantization of \S2 with standard SO(3) invariant
symplectic form of the volume $2\pi R$  (i.e. the algebra $\widehat{Fun(S^2)}$
with $\hbar=1$) is isomorphic to the algebra
$\hat{N}/ D(E)=R$, where $\hat{N}$ is described
in corollary
\ref{cor-commutant}  and
lemma \ref{lem-com-rel}; and $D(E)$ is given by the formula
\ref{hat-e}.
}

 Let us prove that this quantization is the same
as in the previous subsection. The subalgebra generated by $x_i$ with
respect to the star product is isomorphic to $U(so(3))$. Recall that
Casimir element $D(C)$ in $ U(so(3))$ was defined: $D(C):= \sum_{i=1,2,3}
x_i*x_i+1= 4{e}*{f}+ {h}* {h}-2 {h}+1= 4 {f}* {e}+ {h}* {h}+2 {h}+1= 2(
{e}* {f}+ {f}* {e})+ {h}* {h}+1$.

{\Lem $D(C)=D(E)^2$.}

{\bf Proof.} The algebra $\widehat{Fun(\RR^4)}$
 acts by differential operators on the
polynomial algebra $\C[z_1,z_2]$
(where $\bar z_i$ acts as $\partial_{z_i}$ and $z_i$ acts
as multiplication by $z_i$).
The kernel of this action is zero.
Therefore, it suffices to check that the elements $D(C)$ and
$D(E)^2$ act by the same operator. The space of homogeneous polynomials
of degree $n$ is isomorphic to $V_n$ as $sl(2)$-module, and $D(C)$ acts
as the Casimir operator on this space. Thus, according to \ref{cas} for
any homogeneous polynomial $P$ of degree $n$ we have
\bea
D(C)P=(n^2+2n+1)P=(n+1)^2P.
\eea
On the other hand
\bea \label{E-acts}
D(E)^2P=(\frac{1}{2}\sum\limits_{i=1,2}
\partial_{z_i}z_i+z_i\partial_{z_i})^2P
=(1+\sum\limits_{i=1,2}z_i\partial_{z_i})^2P=(n+1)^2P. \eea $\Box$

So we come to the main corollary of this section:

{\Cor \label{cor-iso}
 There is a natural isomorphism
\bea
\hat{N}/ (D(E)=R)\simeq
U(so(3))/ (D(C)=R^2).
\eea}

This proves the desired result that both
quantizations of \S2 are the same.


\subsection{Hilbert space from Hamiltonian reduction.} \label{sect-hil-sp-red}

Let us recall that in the method of Hamiltonian reduction for the
constraints $f_i=0$
the Hilbert space of the reduced system is defined
(according to the conjecture 3)
as the  subspace of the
nonreduced system such that
constraints $D(f_i)  $ acts as zero: $H^{red}=\{v\in H: D(f_i) v=0\}$,
it is clear that the  reduced algebra of functions acts on
this space.

In our case
we have the constraint $E-R=0$, where
$E=D(E)=\frac{1}{2}(p_1^2+q_1^2+p_2^2+q_2^2)$.
The Hilbert space for the quantization
of $\RR^4=\CC^2$ with the standard symplectic structure
$dp_1\wedge dq_1+dp_2 \wedge dq_2$ is
known from any textbook to be $L^2(q_1,q_2)$
with the standard measure $dq_1dq_2$,
another realization for the same space
is the so-called holomorphic realization
$\CC[z_1,z_2]$, (more precisely we should consider
holomorphic functions of $z_1,z_2$ which are square
integrable with the measure $exp(-|z_1|^2-|z_2|^2)$,
but it does not matter for our questions).
In this representation $\bar z_i$ acts as $\partial_{z_i}$.
So we come to:

{\Prop The reduced space for the constraint $E=R$, i.e.
subspace in $\CC[z_1,z_2]$, where $D(E)-R$ acts as zero
is the space of homogeneous polynomials of degree $R-1$, it has the dimension
$R$, the quantization of functions on \S2 acts irreducible and unitary on
this space.}

{\bf Proof} is clear from the formula \ref{E-acts}
and description of representations of $sl(2)$.

{\Cor We obtain that the unique unitary irreducible
representation of $\widehat{(Fun(S^2))}$ can be obtain
by the method of Hamiltonian reduction
so completely confirming the conjecture 3.
Also we obtain the complete agreement of the description
of the Hilbert space from the point of view of Hamiltonian
reduction method and all other points of view:
representation theoretic, geometric quantization with half-forms
and Fedosov's index theorem (see section \ref{sssect-hilb-space1}).}


\section{Comparison with the explicit star product}\label{sec-ex-star}

In this section we recall the explicit
$SO(3)$-invariant star product on \S2 following
\cite{upps} and show that it gives the same quantization
as predicted by our conjectures,
despite that from the first sight we see some
contradiction.

\subsection{ Explicit $SO(3)$-invariant  star product on $\RR^3$.}
Explicit $SO(3)$-invariant star product on $\RR^3$ was found
in \cite{jap} using  earlier work \cite{prej},
later in \cite{aleks} there was proposed invariant star
product on arbitrary coadjoint orbits of semisimple group.
In \cite{upps} it was shown that the last star product
coincides with the one from \cite{jap} in the case of \S2.

Let us recall (by cut and paste from \cite{upps})
 the invariant star product on $\RR^3$ from \cite{jap}. Let $x_i$
$i=1,2,3$ be the coordinates in $\RR^3$,
$r^2=\sum x_i^2$, $\epsilon_{abc}$ - totally antisymmetric
tensor.

\begin{eqnarray}\label{star}
f\star g = fg +\sum_{n=1}^{\infty}C_n(\frac{\hbar}{r})J^{a_1
b_1}\ldots J^{a_n b_n}\partial_{a_1}\ldots
\partial_{a_n}f \partial_{b_1}\ldots \partial_{b_n}g, \nonumber\\
\qquad
\end{eqnarray}
where
\begin{equation}
C_n(\frac{\hbar}{r})=\frac{(\frac{\hbar}{r})^n}{n!(1-\frac{\hbar}{r})
(1-2\frac{\hbar}{r})\cdots(1-(n-1)\frac{\hbar}{r})},
\end{equation}
and
\begin{equation}
J^{ab}=r^2\delta_{ab} - x_ax_b + ir\epsilon_{abc}x_c.
\end{equation}
The star product is defined on $\mathbb{R}^3\backslash
\{\mathbf{0}\}$, but can be restricted to two-spheres centered at
the origin since
because of the property ( see \cite{jap})
$f(r^2)\star g(\mathbf{x})=g(\mathbf{x})\star
f(r^2)=f(r^2)g(\mathbf{x})$ so this property
guarantees, that the ideal generated by $r^2$ is two-sided
and the algebra of functions on \S2 with respect to this
star product is factor by this ideal of the algebra of
functions on $\RR^3$ with respect to the star product above,
and it is rotation invariant since $J^{ab}$ is a covariant 2-tensor.

\subsection{Apparent contradiction of our proposal
and explicit star product calculation}

{\Lem
\bea
[x_a,x_b]_{*}=2 i \hbar \epsilon_{abc}x_c
\\
\sum_{i=1,2,3} x_i*x_i= r^2+2 \hbar r
\eea}

{\bf Proof: } $C_1=\frac{\hbar}{r}$, $~~~~J^{11}=r^2-x_1^2$, $ ~~~~
J^{12}=-x_1x_2+i  r x_3$,
~~~~ so\\
$[x_1,x_2]_{*}= x_1*x_2 -x_2*x_1=
 (x_1x_2 + \frac{\hbar}{r} (-x_1x_2+i  r x_3))
- (x_2x_1 + \frac{\hbar}{r} (-x_2x_1 -i r x_3))
= 2 i \hbar x_3
$
\\
$x_1*x_1=x_1^2+C_1J^{11}= x_1^2 +  \frac{\hbar}{r} (r^2-x_1^2)$, ~~ hence
\\
$\sum_{i=1,2,3} x_i*x_i=  \sum_{i=1,2,3}
x_i^2+ \frac{\hbar}{r} (r^2-x_i^2)=r^2+2\hbar r
$

{\Cor  So we obtain the apparent  contradiction:
as it follows from the lemma above
 the quantization of the \S2 given
by the $\sum_i (x_i^{classical})^2 =R^2$ and the Poisson structure
is $\{x_a^{classical},x_b^{classical} \}=2 \epsilon_{abc} x_c^{classical}$
is the algebra with generators
$x_1,x_2,x_3$ with the relations:
$[x_a,x_b]_{*}= 2 i \hbar \epsilon_{abc}x_c$,
$\sum_{i=1,2,3} x_i*x_i= R^2+2\hbar R$,
but our proposal from the previous sections
predicts that as a quantization of \S2
we should obtain the algebra
with the other answer for the
second relation:
$\sum_{i=1,2,3} x_i*x_i= R^2-1 $
}


\subsection{Solution to the contradiction}

As one can see from the previous corollary
the difference between the two answers is not very big:
if we put
$\hbar=1$, then
our methods of quantization
gives
$\sum_{i=1,2,3} x_i*x_i= R^2-1$ and
star product gives
$\sum_{i=1,2,3} x_i*x_i= R^2+2R=(R+1)^2-1$
 - the same as our method, but with the change $R\to R+1$.

So in order to solve the puzzle we need to explain that star product
\ref{star} quantizes the sphere of radius $R+1$ not  of the radius $R$ as
is seems.

To our luck this essentially has already been done in
\cite{upps} where the characteristic class
of the invariant quantization was found.
Let us mention that it is rather nontrivial
calculation which used the results of Karabegov \cite{Kar2} and
Fedosov-Nest-Tsygan index theorem \cite{Fed,Nest}.

{\Prop \label{char-class} \cite{upps} the characteristic class of the
invariant star product
$\theta = \frac{\omega}{2 \pi \hbar} +\frac{1}{2}\hbar c_1($\S2).
}

Putting $\hbar=1$ we get that $\int_{S^2} \theta = R+1$.
From this we conclude:

{\bf Claim:} the invariant
star product \ref{star} quantizes the symplectic structure on
\S2 which corresponds to sphere with radius $R+1$,
not $R$.

This is more or less by definition of the characteristic class of deformation
quantization. Which measures the difference between the given star product
and the isomorphism class of star products which canonically corresponds
to the given symplectic structure.

It follows from the claim above, that:

{\it We come to complete agreement of our method
of quantization and the explicit star product computation,
due to putting $R-1$ instead of $R$ in the
construction with  the invariant star product quantization
of \S2 we obtain that it has the characteristic class precisely $\omega$
and it gives the same quantum algebra as our methods of quantization.
}


%

%

\subsection{Characteristic classes of deformation quantization}

In this section we briefly recall the material related
to the classification of the star products, in order to
clarify our point of view. (It seem that even among
experts there are some "dark places" in these matters).

%

The naive requirement $f*g-g*f= i \hbar \{ f,g \} mod ~ \hbar^2 $ is not
enough to uniquely define the correspondence between Poisson bracket $\{ ,
\}$ and  the star products. This can be seen from the trivial example: one
can consider zero Poisson bracket $\forall f,g \{ f,g \}=0$ and arbitrary
nonzero Poisson bracket which is multiple of $\hbar$ i.e. $ \{ f,g \}=
\hbar (something)$. The star products for the both of this brackets will
satisfy $f*g-g*f= 0 mod ~ \hbar^2 $.

So  the question  what is the star product corresponding to the given
Poisson bracket arises. And  more generally what is the correspondence
between the brackets and star products. To this question answers the
fundamental theorem of Kontsevich (see section 1.3 in \cite{Kont}). Which
says roughly speaking the following to the given  Poisson bracket one can
construct class of star products, but the algebra of functions with
respect to all these star products are isomorphic. So to the given Poisson
bracket one can construct one algebra up to isomorphism. Moreover
Kontsevich theorem works in the back direction it states that for a given
star product one can describe the class of Poisson brackets (depending on
$\hbar$ in general) deformation quantization of which leads to star
product equivalent to the initial one.

In the case of symplectic manifolds (i.e. when the Poisson bracket is
nondegenerate) the other classification exists:
 star products corresponding to given
symplectic form $\omega$ are classified by $H^2(M)[[\hbar]]$. More
precisely, to the given star product with the property $f*g-g*f=\hbar \{
f,g \}_{\omega} mod \hbar^2 $, one can canonically associate the element
of the affine space $-\frac{\omega}{\hbar}+H^2(M)[[\hbar]]$ (see for
example \cite{Gutt}) and the star-products with the same element from
$H^2(M)[[\hbar]]$ defines the isomorphic algebras. (There is no
contradiction with example described above  and this theorem because in
our example we started with the zero Poisson bracket which is degenerate
so it is not symplectic). The characteristic class is naturally
constructed as a cocycle in the Cech complex of $M$, so it is very hard to
write it down as a de Rham cocycle.

The following conjecture (which states that both classification are
agreed) should be true, but we  do not know the reference:

{\bf Conjecture 4:} {\em If one takes the star product on the symplectic
manifold with the characteristic class $\theta$ then under the bijections
between Poisson brackets  and star products defined by Kontsevich this
star product corresponds to $\theta^{-1}$ }

More precisely one should take nondegenerate representative in the
cohomology class $\theta$, hopefully it can be done.

So this means: if the characteristic class of the quantization is $\theta$
then this quantization really quantizes symplectic manifold with the
symplectic form $\theta$, but not $\omega$.

This clarifies the claim made in previous section.

%


\section{Discussion}

The "weak version" of Conjecture~1 can be formulated as follows. Let
$Z(M)$ be the center of the algebra $\widehat{Fun(M)}$ (by Kotsevich
theorem it is the same as the Poisson center of $Fun(M)$).

{\bf "Weak Conjecture" 1} {\it $Fun(M)$ and $\widehat{Fun(M)}$ are
isomorphic as modules over $Z(M)$}

It is probably a weaker form of our conjecture 1, it can be
explained rather informally, in the following
way: our conjecture says that
 $\widehat{Fun(M)} / D(C-\alpha)$ is quantization of
$Fun(M)/ (C-\alpha)$, where $\alpha$ is arbitrary constant,
so they are isomorphic as modules of constants,
Casimir $C$ here acts as a constant $\alpha$,
due to it is true for all $\alpha$ it should
be true before the factorization.

In \cite{Sh} B. Shoikhet proved that for any Lie algebra $\gg$ there is a
natural isomorphism of $ZU({\gg})$-modules
\bea
S({\gg})/\{S({\gg}),S(\gg)\}\simeq U(\gg)/[U(\gg),U(\gg)].
\eea
The question is
if this isomorphism can be extended to a $ZU(\gg)$-module isomorphism
between $S(\gg)$ and $U(\gg)$.

Let us mention that our conjectures
requires the explicit choice of generators
$f_i$ defining the submanifold.
It is of course not satisfactory,
because the same manifolds can be
defined by different choices of generators.
At the moment the situation
with this question is not clear
for us. It is quite obvious that
linear change of generators
leads to the same quantization.
About the general case it is not quite clear:
hopefully for arbitrary choice of generators $f_i \in I$
 $D(f_i)$ generate the same ideal in $\widehat{Fun(M)}$,
but possibly the ideals generated by them
are different, and there is only  isomorphism
of $\widehat{N_{f_i}}/ \widehat{I_{f_i}}$ for
different of $f_i$.

Moreover the map $D$ is defined not uniquely,
but depends on the auxiliary structures like the choice
of coordinate system in the approach of \cite{Kont},
nevertheless we hope that our conjectures are true
for arbitrary choice of  the map $D$.

Let us mention that the next step to check
our conjectures can be the attempt to
check it for the other coadjoint orbits
of semisimple Lie groups,
for example $gl(n)$.
By the definition they are submanifolds
in $gl(n)$ on the other hand there is the
explicit star product found in \cite{aleks},
and they also can be represented by the hamiltonian
reduction (quiver like description of
coadjoint orbits).
One can hope that all the three approaches
coincide.

It would be also very interesting to find
the generalization of the conjectures above
to the case of infinite-dimensional Lie algebras,
because many interesting spaces like moduli spaces
of flat connections, instantons, etc.
can be obtained by the Hamiltonian reduction
by the action of infinite-dimensional groups.
But in this case seems nothing to be known about
Duflo-Kirillov map, it is believed that it should
be basically the same but there should be some
corrections.

\section{Appendix. General definition of the Duflo-Kirillov-Kontsevich
map}\label{append}

Let us recall the definition of the map $D$. We will follow the letter
from V. Dolgushev to whom we deeply indebted for
the clarifications. One is reffered to \cite{Kont,FTC,CF} for further details.

To any manifold $M$ one can associate two differential graded Lie algebras
(DGLA). The first is the algebra \bea P=\oplus _{k\ge -1}
\Gamma(\wedge^{k+1}TM)\,, \qquad \Gamma(\wedge^0 TM)=Fun(M) \eea of smooth
polyvector fields. The structure of Lie algebra is given by the
Schouten-Nijenhuis bracket $[,]_{SN}$, and the differential is zero.

The second DGLA is the algebra $H$ of polydifferential operators with the
Gerstenhaber bracket  $[,]_G$ and the differential given by \bea
\pa=[m,\bul ]_G, \eea where $m$ is the commutative product $Fun(M)\otimes
Fun(M)\to Fun(M)$.

The solutions of the Maurer-Cartan equation in the first algebra are
Poisson brackets, and the solutions of the Maurer-Cartan equation in the
second algebra are star-products.

The formality quasi-isomorphism of Kontsevich is a (nonlinear)
$L_{\infty}$-morphism $F$ from the DGLA of polyvector fields to the DGLA
of polydifferential operators. The structure maps $F_n$ of this
quasi-isomorphism are described in terms of integrals over configuration
spaces related with the Lobachevsky plane (see \cite{Kont} for more
details). For any solution $\al$ of the  Maurer-Cartan equation in the
DGLA of polyvector fields (i.e. for any Poisson bracket) the
bidifferential operator \bea \label{F} F(\h\al)= \sum_{k=0}^{\infty}
\frac1{k!} F_{k+1} (\h\al,\dots,\, \h\al) \eea satisfies Maurer-Cartan
equation as well.

Furthermore, the formality quasi-isomorphism of Kontsevich gives a
quasi-isomorphism $I$ from the complex of polyvector fields with the
differential \bea d_{\al}=[\h\al,\bul ]_{SN}. \eea to the complex of
polydifferential operators with the differential \bea
\pa_{\al}=[m+F(\h\al),\bul ]_G. \eea This quasi-isomorphism of complexes
is given by the formula:

\bea \label{I} I(\ga)= \sum_{k=0}^{\infty} \frac1{k!} F_{k+1}
(\h\al,\dots,\, \h\al, \ga)\,, \eea where $\ga$ is a cochain in $P$, and
$F_n$ are the structure maps of Kontsevich's quasi-isomorphism. This
quasi-isomorphism is compatible with the cup-product on cohomology of
these complexes.

The fact that a function $C\in Fun(M)$ is central is equivalent to the
fact that $C$ defines a cocycle in $P^{-1}$ with respect to $d_{\al}$, and
the cup-product of such functions is their ordinary product. Analogously,
the fact that a function $C\in \widehat{Fun(M)}$ is central is equivalent
to the fact that $C$ defines a cocycle in $H^{-1}$ with respect to
$\pa_{\al}$, and the cup-product of such functions is their star-product.

Thus, the formula \bea \label{I-C} D(C)= \sum_{k=0}^{\infty} \frac1{k!}
F_{k+1} (\h\al,\dots,\, \h\al, C) \eea defines the desired mapping $D$.

It was proved in \cite{Kont} that it coincides with the map Duflo-Kirillov
map in the case of Lie algebras.

\end{document}